\def\year{2020}\relax
\documentclass[letterpaper]{article} 
\pdfoutput=1
\usepackage{aaai20}  
\usepackage{times}  
\usepackage{helvet} 
\usepackage{courier}  
\usepackage[hyphens]{url}  
\usepackage{graphicx} 
\usepackage{graphicx}  
\title{Multi-Agent Reinforcement Learning as a Computational Tool for Language Evolution Research: Historical Context and Future Challenges }
\author{Cl\'{e}ment Moulin-Frier and Pierre-Yves Oudeyer\\ 
Flowers team, Inria and Ensta ParisTech, France\\ 
clement.moulin-frier@inria.fr, pierre-yves.oudeyer@inria.fr 
}
 
\begin{document}

\maketitle

\begin{abstract}
Computational models of emergent communication in agent populations are currently gaining interest in the machine learning community due to recent advances in Multi-Agent Reinforcement Learning (MARL). Current contributions are however still relatively disconnected from the earlier theoretical and computational literature aiming at understanding how language might have emerged from a prelinguistic substance. The goal of this paper is to position recent MARL contributions within the historical context of language evolution research, as well as to extract from this theoretical and computational background a few challenges for future research. 
\end{abstract}


\section{Origins, formation and forms}

There is a wide variety of approaches to studying the conditions in which human language might have emerged \cite{Christiansen2003}. As we will see, computer simulations have historically played an important role in the field. We can divide the problem in three sub-parts \cite{oudeyer:hal-00818204}. Firstly, the study of the \textbf{forms} of language, i.e. of the structure of the phonemic, semantic, syntactic or pragmatic systems constituting it. Secondly, the study of its \textbf{formation}, i.e. of the genesis of these forms through sensory-motor, cognitive, environmental, social, cultural or evolutionary processes. Thirdly, the study of the \textbf{origins}, i.e. of the biological and environmental conditions that could have bootstrapped the formation process. 

Under the infinite variety of its \textbf{forms}, human language is characterized by obvious regularities, the universals of language, which we find for example at the phonemic level (with vowels present in almost all languages of the world, \cite{Maddieson1989}) and syntactic level (all languages have a recursive hierarchical structure, see e.g. \cite{pinker1990natural}). A fundamental research question concerns the \textbf{origins} of these regularities. Three main arguments are proposed in the literature. In the Chomskyan view of a genetically specified language acquisition device \cite{Chomsky1965}, a common innate language competence shared by all humans would explain the regularities observed in the different languages. Another view about the universal properties of human languages may be found in the hypothesis of a common origin, by which human languages would derive from an African mother tongue \cite{Ruhlen1996}, imposing some common traces in spite of further cultural evolution producing their diversity. A third view considers that the forms of human language are the emergent product of an optimization process, inducing some commonality in the achieved solutions because of commonality in the cognitive mechanisms at hand, and because of common exterior constraints. This is the view first popularized by \cite{lindblom1984}, through a proposal to \textit{"derive language from non-language"}. This last proposal opened a whole research program aiming at understanding the \textbf{formation} of human language, i.e. how a non-linguistic substance consisting in all the biological, cognitive and environmental mechanisms present before language, could both bootstrap its emergence and shape its universal properties, its form. 

\section{Theories on the formation of language}

A large proportion of these theories postulate of a joint evolution of cooperative and communicative behaviors \cite{Smith2010,gardenfors2002cooperation,Ghazanfar2014,Tomasello2012}. It is in particular the central thesis of the theory developed by Michael Tomasello, who proposes that \textit{"humans' species-unique forms of cooperation --as well as their species-unique forms of cognition, communication, and social life—all derive from mutualistic collaboration (with social selection against cheaters)"} \cite{Tomasello2012} . In this view, it is the constraints imposed by the ecological niche occupied by human beings that has forced them to jointly develop complex collaborative and communicative behaviors, in a context of interdependence requiring the sharing of intentions. We also find compatible arguments in the mirror system hypothesis developed by Michael Arbib \cite{arbib_monkey-like_2005} proposing that language evolution is grounded in the sensory-motor integration required for the execution and the observation of transitive actions towards objects, enabling other's intention recognition and providing the bases of a syntactic structure \cite{Roy2005} (see also \cite{Iriki2012} for theoretical propositions on the coevolution of tool use and language in humans). Finally, the social complexity hypothesis suggests that groups with complex social structures require more complex communication systems to regulate interactions between group members \cite{Freeberg2012}.

Other theories highlight the role of sensory-motor learning and exploration as a key element to understand how speech communication could emerge from pre-existing morphological, perceptual and behavioral constraints \cite{lindblom1984,macneilage98,Schwartz2012a}. A few theoretical contributions have proposed a potential role of curiosity-driven exploration in both language acquisition \cite{Oller2000} and evolution \cite{oudeyer_smith_evo_devo}.  

\section{From verbal to computational descriptions}

A major limitation of most of the theories mentioned above is that they are described in a verbal form. They are of course supported by experimental data but the description of the underlying hypotheses regarding the formation of linguistic structures mostly relies on a verbal explanation. This can be problematic because the aim of those theories is precisely to describe a complex dynamical process where linguistic structures emerge from multiple constraints in a prelinguistic environment (e.g. morphological, sensory-motor, cognitive, developmental, evolutionary or cultural constraints). Computer simulation is required to study the emergent properties of such a complex dynamical system.

For this reason, computational modeling has played a major role in language evolution research. Already in the 70s, Lindblom's "Dispersion Theory" \cite{Liljencrants1972} proposed that human phonological systems are optimized for maximizing auditory distances between phoneme pairs in order to enhance distinguishability. In these early contributions, language forms (e.g. the form of vowel systems) are considered as the equilibrium of a macroscopic system, analog to how thermodynamics describes changes in macroscopic physical quantities. In the 90s, these "global" approaches were completed by "local" approaches, were the equilibrium emerges from the interaction of "microscopic" elements, analog to statistical mechanics showing how the concepts from macroscopic observations are related to the description of microscopic states. These local approaches usually involve interacting prelinguistic agents and study how properties of human language can emerge from these interactions. A well-known example is the naming game paradigm showing how a shared communication system, associating signals emitted by the agents with semantic references to the external world, can self-organize out of a decentralized learning process from the local interactions between the agents \cite{Steels97} (see \cite{deBoer2000,moulinfrier2015cosmo} for extensions to vocal communication and \cite{oudeyer2005self,Boer2010} for extensions to combinatorial communication). 
However, these naming game models rarely address the issue of the functionality of communication (i.e. why to communicate?). Models from the field of evolutionary robotics \cite{quinn2001evolving,Grouchy2016} have the advantage of considering more realistic interaction scenarios than naming games but they specifically focus on genetic evolution algorithms, which do not consider the role of sensory-motor learning processes. 

Computational models of emergent communication in agent populations are currently gaining interest in the machine learning community, due in particular to recent advances in Multi-Agent Reinforcement Learning (MARL) (see \cite{hernandez2019survey} for a survey). These new possibilities have allowed to overcome certain limitations of earlier contributions in two main directions. On the one hand, the paradigm of naming games presented above has been extended to more realistic references to the external world, learning directly from observations of raw images \cite{Lazaridou2018}. On the other hand, recent contributions based on the paradigm of partially-observable cooperative Markov games \cite{Littman1994,Leibo2017c} have shown how a communication system can emerge to solve cooperative tasks in sequential environments \cite{Sukhbaatar2016,Mordatch2017,foerster2016learning}. These contributions adopt an utilitarian view of communication, where communication emerges as a way to solve complex cooperative tasks \cite{Gauthier2016}.

\section{Extracting future challenges for MARL}

The utilitarian approach relying on partially observable cooperative Markov games provides a powerful conceptual and computational framework for modeling emergent communication as a way to solve complex problems in sequential environments. However, existing contributions are still relatively disconnected from the earlier literature presented in the previous section. In this section, we will extract from this theoretical and computational background a few challenges for future MARL research.

\subsection{Decentralized learning}
As mentioned in the previous section, the first models attempting to predict language forms from a prelinguistic substance adopted a global, macroscopic approach. This global approach has then be complemented by a local, microscopic approach where language forms emerge from the repeated interactions between individual agents.

A large proportion of current MARL contributions rely on \textit{centralized learning decentralized execution} algorithms \cite{Sukhbaatar2016,Mordatch2017,foerster2016learning}, analog to a global macroscopic approach. While centralized learning is able to efficiently solve complex problems, the lack of biological plausibility strongly limits its use in language evolution research. Contributions relying on decentralized learning \cite{anonymous2019intrinsic} are less efficient from a performance point of view but have the advantage of highlighting important issues regarding the unstable nature of cooperative and communicative behavior in multi-agent settings, due e.g. to the non-stationarity it induces. Solving such issues is an important challenge in both MARL and language evolution research.

\subsection{Role of morphological and sensory-motor constraints}
Current MARL contributions mostly rely on an idealized communication channel where the signal produced by an agent is directly broadcasted to other agents \cite{Sukhbaatar2016,Mordatch2017,foerster2016learning}, similar to earlier contributions based on the naming game paradigm. In contrast, speech communication is strongly shaped by sensory-motor constraints, involving the control of vocal articulators (e.g. the jaw, the tongue, the lips) for modulating a sound wave resulting in the perception of acoustic features. Vocal control is actually a classical robotic problem, where the agent has to decide how to move vocal articulators to reach acoustic targets. This control problem is a difficult one due to the complex morphology of the vocal tract, the highly non-linear nature of the articulatory-to-acoustic transformation, as well as the presence of acoustic noise in the environment. Earlier contributions have studied how vocal communication can emerge from the interaction of sensory-motor agents equipped with articulatory synthesizers, i.e. computer models of the human vocal tract able to generate sound waves from articulator trajectories \cite{moulinfrier2015cosmo,Moulin-Frier_Frontiers_2013}. This resulted in multi-agent simulations able to predict the statistical tendencies of the phonological systems used in world languages \cite{Oudeyer2005}, as well as to test hypotheses regarding the influence of prelinguistic orofacial behaviors on the syllabic structure of speech communication (\cite{moulinfrier2015cosmo}, following an hypothesis from \cite{macneilage98}). Introducing biologically plausible sensory-motor abilities of signal production and perception in MARL models would allow to extend the aforementioned results to more complex environments and learning abilities.

\subsection{Role of intrinsic motivation}
A few theoretical contributions have proposed a potential role of curiosity-driven exploration in both language acquisition \cite{Oller2000} and evolution \cite{oudeyer_smith_evo_devo}. Active exploration can spontaneously generate diverse behaviors from modality-independent and task-independent internal drives. Such spontaneous behavior can result in vocal activity that may have bootstrapped the emergence of communication. This hypothesis is supported by computational simulations showing a role of curiosity-driven exploration in vocal development \cite{Moulin-Frier_Frontiers_2013}, social affordance discovery \cite{Oudeyer2006} and the active control of complexity growth in naming games \cite{schueller2015active}.

Despite recent progress in curiosity-driven RL \cite{pathak2017curiosity,colas2018curious}, very few MARL contributions have used such algorithms for studying emergent communication (see \cite{anonymous2019intrinsic} but which is specific to social interactions on a single task). It is a promising direction of research to explore how general-purpose curiosity-driven multi-task reinforcement learning algorithms \cite{colas2018curious} can be integrated in multi-agent environments to encourage the discovery of complex communication systems supporting the acquisition of an open-ended repertoire of cooperative skills. A key step in this direction has recently been proposed in the IMAGINE architecture~\cite{colas2020LanguageCognitiveTool}, where an agent uses language compositionality to generate new goals by composing known ones.

\subsection{Emergent complexity}
Earlier contributions in language evolution modeling has often been limited by the use of simplistic environments and learning abilities. Recent advances in MARL can allow to overcome these limitations to show how language complexity can emerge as a way to optimize behavior in complex cooperative environments. In particular, recent contributions in MARL have shown how an autocurriculum of increasingly complex behaviors can emerge from agent's coadaptation in mixed cooperative-competitive environments \cite{bansal2018emergent,Baker2019}. Can such an auto-curriculum through coadaptation favor the emergence of increasingly complex communicative systems? In turn, can complex communication favor the emergence of increasingly complex cooperative strategies? Addressing these open questions can potentially help to understand the processes that have shaped the impressive complexity of human language.

\section{Conclusion}

Recent advances in MARL provides a powerful conceptual and computational framework for modeling emergent communication as a way to solve complex problems in sequential environments. There are however important differences in the methodology and the objectives between 1) implementing efficient and robust multi-agent systems learning how to communicate for solving complex problems (as it is the case in the majority of recent MARL contributions), vs. 2) using multi-agent learning as a computational tool for better understanding human language evolution (an approach which has historically played an important role in language evolution research, see \cite{oudeyer:hal-00818204} for an epistemological analysis). In this paper we have reviewed earlier computational contributions and have extracted from them a few future challenges for MARL research.

\fontsize{9.0pt}{10.0pt} \selectfont
\bibliographystyle{aaai}
\bibliography{../../../../readings/bibtex/library.bib}

\end{document}